\documentclass[a4paper,12pt]{article}

\usepackage{hyperref}
\hypersetup{
    colorlinks=true,
    linkcolor=blue,
    filecolor=magenta,      
    urlcolor=cyan,
    citecolor=green
    }

\usepackage{latexsym}
\usepackage{a4wide}
\usepackage{graphicx, subfigure} 
\usepackage{cite} 
\usepackage{tabularx}
\usepackage{array}
\usepackage{graphics}     
\usepackage{amsmath}
\usepackage{amssymb}
\usepackage{longtable} 
\usepackage{verbatim} 
\usepackage[table]{xcolor}
\usepackage{booktabs}
\usepackage{hyperref} 
\usepackage[utf8]{inputenc}
\usepackage{multirow}
\newcolumntype{C}[1]{>{\centering\arraybackslash}p{#1}}
\newcolumntype{L}[1]{>{\raggedright\arraybackslash}p{#1}}
\newcolumntype{R}[1]{>{\raggedleft\arraybackslash}p{#1}}

\begin{document}

\hfill {\tt CERN-TH-2022-166, MITP-22-081}  

\def\thefootnote{\fnsymbol{footnote}}
 
\begin{center}

\vspace{2.cm}

{\Large\bf {Neutral current $B$-decay anomalies}}

\setlength{\textwidth}{11cm}
                    
\vspace{2.cm}

{\large\bf  
T.~Hurth$^{a}$,\,
F.~Mahmoudi$^{b,c}$,\,
D.~Mart\'inez Santos$^{d}$,\,
S.~Neshatpour$^{e}$
}
 
\vspace{0.5cm}
{\em $^a$PRISMA+ Cluster of Excellence and  Institute for Physics (THEP),\\
Johannes Gutenberg University, D-55099 Mainz, Germany}\\[0.2cm]
{\em $^b$Universit\'e de Lyon, Universit\'e Claude Bernard Lyon 1, CNRS/IN2P3, \\
  Institut de Physique des 2 Infinis de Lyon, UMR 5822, F-69622, Villeurbanne, France}\\[0.2cm]
{\em $^c$CERN, Theoretical Physics Department, CH-1211 Geneva 23, Switzerland,}\\[0.2cm]
{\em $^d$Instituto Galego de F\'isica de Altas Enerx\'ias,\\ Universidade de Santiago de Compostela, Spain}\\[0.2cm]
{\em $^e$INFN-Sezione di Napoli, Complesso Universitario di Monte S. Angelo,\\ Via Cintia Edificio 6, 80126 Napoli, Italy}

\end{center}

\renewcommand{\thefootnote}{\arabic{footnote}}
\setcounter{footnote}{0}

\vspace{1.cm}
\thispagestyle{empty}
\centerline{\bf ABSTRACT}
\vspace{0.5cm}
We discuss the implications of $b \to s \ell^+\ell^-$ measurements and their deviations with respect to the Standard Model predictions in a model-independent framework. We highlight in particular the impact of the recent updated measurements including the updated $B_s \to \phi \mu^+\mu^-$ branching ratios and angular observables, the recent CMS measurement of the branching ratio of $B_s\to\mu^+ \mu^-$, and the LHCb measured lepton flavour universality violating ratios $R_{K_S^0}$ and $R_{K^{*+}}$.  
In addition, we check the compatibility of the new physics effect for the theoretically clean observables with the rest of the neutral $B$ decays observables.

\newpage

\section{Introduction}
In the last few years, since the measured deviation in the angular observable $P_5^\prime$ of the $B \to K^* \mu^+ \mu^-$ decay~\cite{LHCb:2013ghj}, there have been several measurements in neutral $B$-decays indicating tension with the Standard Model (SM). 
Updated measurements by LHCb for the $P_5^\prime(B \to K^* \mu^+ \mu^-)$ have persistently shown tension with the SM which can be explained with short distance new physics (NP) contributions~\cite{LHCb:2015svh,LHCb:2020lmf}. This is also the case of the overall $B \to K^* \mu^+ \mu^-$ angular observables and is supported in addition (see e.g.~\cite{Hurth:2020ehu}) by the angular analysis of its isospin partner in the recent measurement of $B^+ \to K^{*+} \mu^+ \mu^-$~\cite{LHCb:2020gog}. The $B_s \to \phi \mu^+ \mu^-$ branching fraction~\cite{LHCb:2015wdu,LHCb:2021xxq,LHCb:2021zwz} also indicates tensions with the SM and is measured to be below the SM prediction. This trend is seen in several other $b\to s \ell^+ \ell^-$ branching fractions such as $B \to K \mu^+ \mu^-$~\cite{LHCb:2014cxe} and $\Lambda_b \to \Lambda \mu^+ \mu^-$~\cite{LHCb:2015tgy}. 
Since the branching fractions are dependent on the relevant local form factors, they suffer from large theoretical uncertainties. In contrast, the angular observables have a reduced sensitivity to the form factor uncertainties, but they are still dependent on the non-local hadronic contributions whose size are not fully known in QCD factorisation. Consequently the significance of the anomalies are dependent on the estimated size of the non-local effects.
Recent theoretical progress for a better control of these effects can be found in Refs.~\cite{Bobeth:2017vxj,Gubernari:2020eft,Gubernari:2022hxn}.

A set of observables to test lepton flavour universality violation (LFUV) in $b\to s \ell^+ \ell^-$ transitions is defined as $R_H = (B\to H \mu^+\mu^-)/(B\to H e^+e^-)$ with $H=K^+,K^*,\phi,...$~\cite{Hiller:2003js}.
Unlike the observables mentioned in the previous paragraph, these ratios are very precisely known in the SM.
There have been signs of deviation from the SM in the LFUV ratios for the case of $R_K$~\cite{LHCb:2014vgu,LHCb:2019hip,LHCb:2021trn} and $R_{K^*}$~\cite{LHCb:2017avl}.
The recent measurements of $R_{K_S^0}$ and $R_{K^{*+}}$~\cite{LHCb:2021lvy} although within 2$\sigma$ of the SM prediction, show the same trend as their isospin partners with the central values below the SM predictions.
Incidentally, there have also been a slight sign of LFUV in flavour changing neutral current processes in the Kaon sector~\cite{DAmbrosio:2022kvb} (currently the experimental uncertainty is quite large for these processes).

The significance of each of the $B$-anomalies, individually is around $\sim2-3\sigma$, however collectively they can be explained by common NP scenarios and have a much larger significance in a global analysis~\cite{Hurth:2021nsi,Alguero:2021anc,Altmannshofer:2021qrr,Ciuchini:2020gvn,Geng:2021nhg,Mahmoudi:2022hzx}.

\begin{figure}[t!]
\centering
\includegraphics[width=0.55\textwidth]{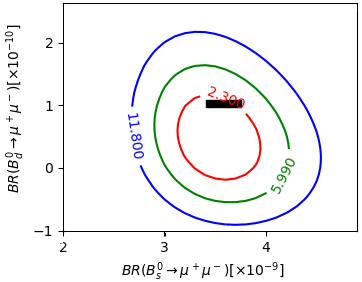}\\
\vspace{-0.1cm}
\caption{Two dimensional likelihood plot of BR($B_{s,d}\to \mu^+\mu^-$).} 
\label{fig:Bsmumu}
\end{figure}

Another precisely predicted observable with an uncertainty of less than 5\% in its SM prediction is BR($B_s \to \mu^+ \mu^-$) which has been measured by several experiments. Previous  
measurements of ATLAS~\cite{ATLAS:2018cur}, CMS~\cite{CMS:2019bbr} and LHCb~\cite{LHCb:2021awg,LHCb:2021vsc} were in about $2\sigma$ tension with the SM prediction~\cite{Hurth:2021nsi}. However, the situation has changed with the recent data from CMS~\cite{CMS:2022dbz}. For our fits, we combined the ATLAS~\cite{ATLAS:2018cur} and LHCb~\cite{LHCb:2021awg,LHCb:2021vsc} results together with the recent CMS~\cite{CMS:2022dbz} measurement, considering a joint 2D likelihood as shown in Fig~\ref{fig:Bsmumu}. We obtained the experimental combined value of the $B_s \to \mu^+ \mu^-$ branching ratio to be
\begin{equation}\label{eq:BsmumuComb}
\rm{BR}(B_s \to \mu^+ \mu^-)_{\rm{exp}}^{\rm{comb.}} = \left(3.52_{-0.30}^{+0.32}\right)\times 10^{-9}\,,
 \end{equation}
which is within $1\sigma$ of the SM prediction.

\section{Coherence of clean observables with the rest of the rare \emph{B}-decay observables}
In order to examine the consistency of the implication of the clean observables for new physics as compared to the rest of the observables~\cite{Hurth:2017hxg,Arbey:2019duh}, we perform two sets of fits; one to the clean observables where we consider $R_K, R_{K^*}$ as well as their isospin partners $R_{K_S^0}$ and $R_{K^{*+}}$~\cite{LHCb:2021lvy} and also BR($B_{s,d} \to \mu^+ \mu^-$), and another one considering the rest of the $bs\ell\ell$ observables.  
The observable calculations and the $\chi^2$ fitting is done using the SuperIso public program~\cite{Mahmoudi:2007vz,Mahmoudi:2008tp,Mahmoudi:2009zz,Neshatpour:2021nbn,Neshatpour:2022fak}. 

\subsection{Clean observables}\label{sec:clean}
In Table~\ref{tab:1D_Full2022_clean} we give the one-dimensional NP fits to clean observables and compare them with our 2021 fit results~\cite{Hurth:2021nsi}.
Compared to Ref.~\cite{Hurth:2021nsi}, we now include the two LFUV ratios $R_{K_S^0}$ and $R_{K^{*+}}$~\cite{LHCb:2021lvy} as well as the $R_K$ measurement by Belle~\cite{BELLE:2019xld} in the [1,6] GeV$^2$ bin and the updated combination for BR($B_s \to \mu^+ \mu^-$) as given in Eq.~(\ref{eq:BsmumuComb}). 
\begin{table}[h!]
\begin{center}
\setlength\extrarowheight{5pt}
\scalebox{0.75}{
\begin{tabular}{|C{1.2cm}|C{2.7cm}|C{1.7cm}|C{1.7cm}|}
\hline 
 \multicolumn{4}{|c|}{\footnotesize Only LFUV ratios and $B_{s,d}\to \ell^+ \ell^-$ \vspace{-0.1cm}} \\
 \multicolumn{4}{|c|}{{\bf 2021 fit results}\quad ($\chi^2_{\rm SM}=28.19$)} \\ \hline
    & b.f. value & $\chi^2_{\rm min}$ & ${\rm Pull}_{\rm SM}$  \\ 
\hline \hline
$\delta C_{9} $    	& $ 	-1.00	\pm	6.00	 $ & $ 	28.1	 $ & $	0.2	\sigma	 $  \\
$\delta C_{9}^{e} $    	& $ 	0.80	\pm	0.21	 $ & $ 	11.2	 $ & $	4.1	\sigma	 $  \\
$\delta C_{9}^{\mu} $    	& $ 	-0.77	\pm	0.21	 $ & $ 	11.9	 $ & $	4.0	\sigma	 $  \\
\hline										
$\delta C_{10} $    	& $ 	0.43	\pm	0.24	 $ & $ 	24.6	 $ & $	1.9	\sigma	 $  \\
$\delta C_{10}^{e} $    	& $ 	-0.78	\pm	0.20	 $ & $ 	9.5	 $ & $	4.3	\sigma	 $  \\
$\delta C_{10}^{\mu} $    	& $ 	0.64	\pm	0.15	 $ & $ 	7.3	 $ & $	4.6	\sigma	 $  \\
\hline							          			
\hline
$\delta C_{\rm LL}^e$	& $ 	0.41	\pm	0.11	 $ & $ 	10.3	 $ & $	4.2	\sigma	 $  \\
$\delta C_{\rm LL}^\mu$ 	& $ 	-0.38	\pm	0.09	 $ & $ 	7.1	 $ & $	4.6	\sigma	 $  \\
\hline
\end{tabular}
\qquad
\begin{tabular}{|C{1.2cm}|C{2.7cm}|C{1.7cm}|C{1.7cm}|}
\hline 
 \multicolumn{4}{|c|}{\footnotesize Only LFUV ratios and $B_{s,d}\to \ell^+ \ell^-$ \vspace{-0.1cm}} \\
 \multicolumn{4}{|c|}{{\bf 2022 fit results}\quad ($\chi^2_{\rm SM}=30.63$)} \\ \hline
    & b.f. value & $\chi^2_{\rm min}$ & ${\rm Pull}_{\rm SM}$  \\ 
\hline \hline
$\delta C_{9} $    	& $ 	-2.00	\pm	5.00	 $ & $ 	30.5	 $ & $	0.4	\sigma	 $  \\
$\delta C_{9}^{e} $    	& $ 	0.83	\pm	0.21	 $ & $ 	10.8	 $ & $	4.4	\sigma	 $  \\
$\delta C_{9}^{\mu} $    	& $ 	-0.80	\pm	0.21	 $ & $ 	11.8	 $ & $	4.3	\sigma	 $  \\
\hline										
$\delta C_{10} $    	& $ 	0.03	\pm	0.20	 $ & $ 	30.6	 $ & $	0.1	\sigma	 $  \\
$\delta C_{10}^{e} $    	& $ 	-0.81	\pm	0.19	 $ & $ 	8.7	 $ & $	4.7	\sigma	 $  \\
$\delta C_{10}^{\mu} $    	& $ 	0.50	\pm	0.14	 $ & $ 	16.2	 $ & $	3.8	\sigma	 $  \\
\hline							          			
\hline										
$\delta C_{\rm LL}^e$	& $ 	0.43	\pm	0.11	 $ & $ 	9.7	 $ & $	4.6	\sigma	 $  \\
$\delta C_{\rm LL}^\mu$    	& $ 	-0.33	\pm	0.08	 $ & $ 	12.4	 $ & $	4.3	\sigma	 $  \\
\hline
\end{tabular}
} 
\caption{Comparison of the fits to clean observables with the 2021 fit results~\cite{Hurth:2021nsi} on the left and the updated 2022 fits on the right.}
\label{tab:1D_Full2022_clean} 
\end{center} 
\end{table}
\begin{figure}[ht!]
\centering
\includegraphics[width=0.49\textwidth]{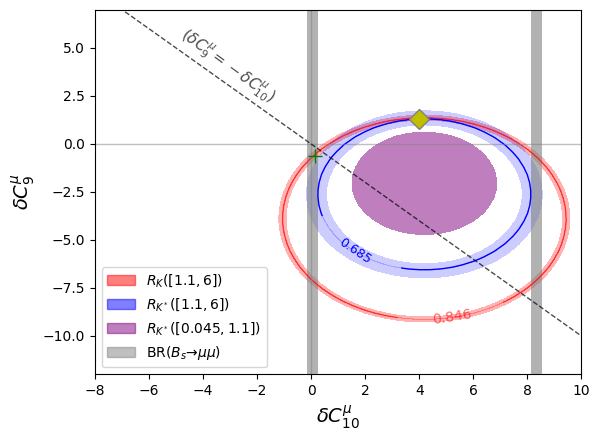}
\includegraphics[width=0.49\textwidth]{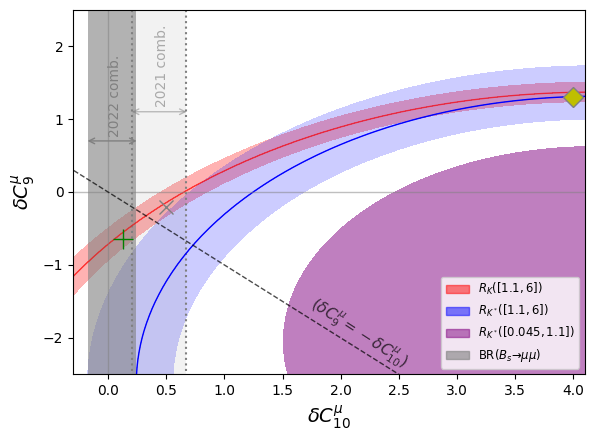}
\vspace{-0.3cm}
\caption{The prediction of $R_{K^{(*)}}$ and BR($B_s\to\mu^+\mu^-$) within $1\sigma \left(=\sqrt{\sigma_{\rm th}^2 + \sigma_{\rm exp}^2}\right)$ of their measured values. On the right plot we have the zoomed-in version of the left plot.  The dark gray band indicates the $1\sigma$ region corresponding to the updated combination of BR($B_s \to \mu\mu$) and the lighter gray region (on the right plot) with the dotted borders corresponds to the 2021 combination. 
The yellow diamond indicates the best fit value to  $R_{K^{(*)}}$, while the green plus sign (gray cross) corresponds to the best fit point when the 2022 (2021) combination for BR($B_s \to \mu\mu$) is included in the fit.} 
\label{fig:BsmumuRole}
\end{figure}
While the significance of NP in $C_9^{e,\mu}$ or $C_{10}^e$ has slightly increased, the $C_{10}^\mu$ solution is now less favoured compared to the 2021 results~\cite{Hurth:2021nsi}. This is expected as the new combination of BR($B_s \to \mu^+\mu^-$) is now in much better agreement with the SM prediction and constrains more $C_{10}^\mu$.
The inclusion of BR($B_s \to \mu^+ \mu^-$) in this set of observables is crucial in breaking the degeneracy between NP in $\delta C_9^\mu$ and $\delta C_{10}^\mu$ for explaining the measured values of the LFUV ratios as can be clearly seen in Fig.~\ref{fig:BsmumuRole} where without BR($B_s \to \mu^+ \mu^-$) the best fit point of $R_{K^{(*)}}$ is given by the yellow diamond while including it moves the best fit value to the green plus sign. The impact of the updated value of the BR($B_s \to \mu^+ \mu^-$) can bee seen in the right plot by comparing the green plus sign with the gray cross corresponding to the best fit point when the 2021 combination for BR($B_s \to \mu^+ \mu^-$) was considered.

\begin{table}[t]
\begin{center}
\setlength\extrarowheight{5pt}
\scalebox{0.75}{
\begin{tabular}{|C{1.2cm}|C{2.7cm}|C{1.7cm}|C{1.7cm}|}
\hline 
 \multicolumn{4}{|c|}{\footnotesize All observables except LFUV ratios and $B_{s,d}\to \ell^+ \ell^-$ \vspace{-0.1cm}} \\ 
 \multicolumn{4}{|c|}{{\bf 2021 fit results}\quad ($\chi^2_{\rm SM}=200.1$)} \\ \hline
                          & b.f. value & $\chi^2_{\rm min}$ & ${\rm Pull}_{\rm SM}$  \\ 
\hline \hline
$\delta C_{9} $    	& $ 	-1.01	\pm	0.13	 $ & $ 	158.2	 $ & $	6.5	\sigma	 $  \\
$\delta C_{9}^{e} $    	& $ 	0.70	\pm	0.60	 $ & $ 	198.8	 $ & $	1.1	\sigma	 $  \\
$\delta C_{9}^{\mu} $    	& $ 	-1.03	\pm	0.13	 $ & $ 	156.0	 $ & $	6.6	\sigma	 $  \\
\hline										
$\delta C_{10} $    	& $ 	0.34	\pm	0.23	 $ & $ 	197.7	 $ & $	1.5	\sigma	 $  \\
$\delta C_{10}^{e} $    	& $ 	-0.50	\pm	0.50	 $ & $ 	199.0	 $ & $	1.0	\sigma	 $  \\
$\delta C_{10}^{\mu} $    	& $ 	0.41	\pm	0.23	 $ & $ 	196.5	 $ & $	1.9	\sigma	 $  \\
\hline							          			
\hline
$\delta C_{\rm LL}^e$	& $ 	0.33	\pm	0.29	 $ & $ 	198.9	 $ & $	1.1	\sigma	 $  \\
$\delta C_{\rm LL}^\mu$	& $ 	-0.75	\pm	0.13	 $ & $ 	167.9	 $ & $	5.7	\sigma	 $  \\
\hline										
\end{tabular}
\qquad
\begin{tabular}{|C{1.2cm}|C{2.7cm}|C{1.7cm}|C{1.7cm}|}
\hline 
 \multicolumn{4}{|c|}{\footnotesize All observables except LFUV ratios and $B_{s,d}\to \ell^+ \ell^-$ \vspace{-0.1cm}} \\ 
 \multicolumn{4}{|c|}{{\bf 2022 fit results}\quad ($\chi^2_{\rm SM}=221.8$)} \\ \hline
                          & b.f. value & $\chi^2_{\rm min}$ & ${\rm Pull}_{\rm SM}$  \\ 
\hline \hline
$\delta C_{9} $    	& $ 	-0.95	\pm	0.13	 $ & $ 	185.1	 $ & $	6.1	\sigma	 $  \\
$\delta C_{9}^{e} $    	& $ 	0.70	\pm	0.60	 $ & $ 	220.5	 $ & $	1.1	\sigma	 $  \\
$\delta C_{9}^{\mu} $    	& $ 	-0.96	\pm	0.13	 $ & $ 	182.8	 $ & $	6.2	\sigma	 $  \\
\hline										
$\delta C_{10} $    	& $ 	0.29	\pm	0.21	 $ & $ 	219.8	 $ & $	1.4	\sigma	 $  \\
$\delta C_{10}^{e} $    	& $ 	-0.60	\pm	0.50	 $ & $ 	220.6	 $ & $	1.1	\sigma	 $  \\
$\delta C_{10}^{\mu} $    	& $ 	0.35	\pm	0.20	 $ & $ 	218.7	 $ & $	1.8	\sigma	 $  \\
\hline							          			
\hline
$\delta C_{\rm LL}^e$	& $ 	0.34	\pm	0.29	 $ & $ 	220.6	 $ & $	1.1	\sigma	 $  \\
$\delta C_{\rm LL}^\mu$    	& $ 	-0.64	\pm	0.13	 $ & $ 	195.0	 $ & $	5.2	\sigma	 $  \\
\hline										
\end{tabular}
} 
\caption{Comparison of the fits to all observables except the clean ones with the 2021 fit results on the left and the updated 2022 fits on the right.}
\label{tab:1D_Full2022_rest} 
\end{center} 
\end{table}
\subsection{All except the clean observables}\label{sec:rest}
We consider now the 1-dimensional NP fits to the rest of the observables, excluding the LFUV ratios and $B_{s,d}\to \ell^+ \ell^-$. We assume 10\% power correction for the non-factorisable contributions beyond QDC factorisation~\cite{Hurth:2020rzx,Chobanova:2017ghn,Neshatpour:2017qvi}. 
Compared to Ref.~\cite{Hurth:2021nsi} we use the updated LHCb results for the $B_s \to \phi \mu^+\mu^-$ observables~\cite{LHCb:2021xxq,LHCb:2021zwz} with 8.4 fb$^{-1}$ of data. The CMS measurement for $F_H(B^+\to K^+\mu^+\mu^-)$~\cite{CMS:2018qih} and the LHCb measurement of the angular observables of $B\to K^* e^+ e^-$~\cite{LHCb:2020dof} have also been considered.
As can be seen in Table~\ref{tab:1D_Full2022_rest}, the hierarchy of the preferred NP contributions is similar to the 2021 results, where the most preferred scenarios are still NP in lepton flavour violating $\delta C_9^\mu$ and NP in lepton flavour universal $\delta C_9$ with the third most preferred description given by NP in the chiral basis $\delta C_{\rm LL}^\mu$.
The above mentioned scenarios however are showing a $\sim 0.4\sigma$ reduced significance compared to our 2021 results which is mainly due to the updated $B_s \to \phi \mu^+ \mu^-$ experimental data.

For the fit to all observables except the clean ones there is no significant indication for NP within the electron sector since not only the measurements in the electron sector are in good agreement with their SM predictions, there are also far less data compared to the decays with muons.
Comparing the result of Table~\ref{tab:1D_Full2022_rest} with the result of the previous subsection (Table~\ref{tab:1D_Full2022_clean}) we see that there is not a full agreement for the preferred scenarios, however, there are common scenarios such as NP contributions to $\delta C_9^\mu$~\cite{Hurth:2016fbr,Hurth:2014vma} which have a large significance for both datasets with best fit points that agree within $1\sigma$. 

\begin{figure}[h!]
\centering
\includegraphics[width=0.48\textwidth]{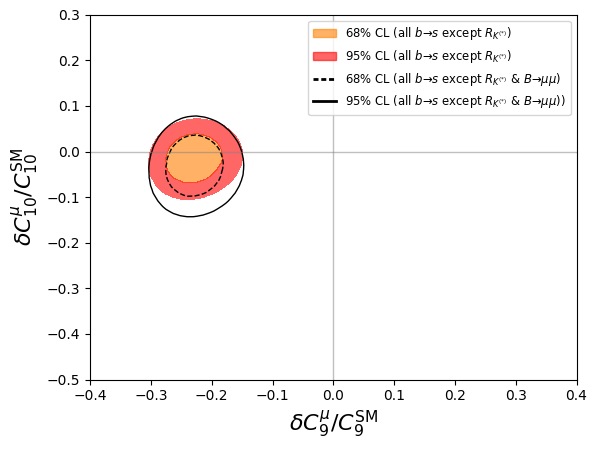}\quad
\includegraphics[width=0.48\textwidth]{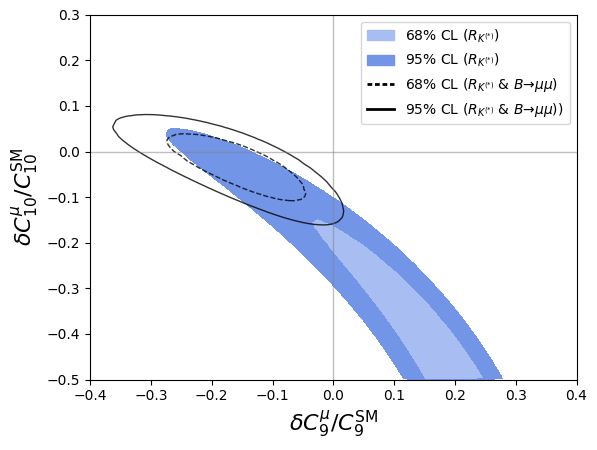}\quad
\vspace{-1.cm}
\caption{Two-dimensional fits to the clean observables on the right and to the rest of the observables on the left. 
} 
\label{fig:clean_vs_rest}
\end{figure}

The compatibility of the two-dimensional NP fits to ``clean observables'' and the NP fits to ``all observables except the clean ones'' can be seen in Fig.~\ref{fig:clean_vs_rest} where also the significant impact of including or removing BR($B_s \to \mu^+\mu^-$) from each dataset is clearly visible, especially for the clean observables.

\section{Global fits to all \texorpdfstring{$b \to s \ell^+ \ell^-$}{b->sll} observables}
For a global analysis of the NP implications of rare $B$-decays, we need to take into account all relevant $b\to s$ decays combining the datasets of section~\ref{sec:clean} and section~\ref{sec:rest}. We assume 10\% error for the power corrections when applicable.

\subsection{One- and two-dimensional fits}

\begin{figure}[ht!]
\centering
\includegraphics[width=0.49\textwidth]{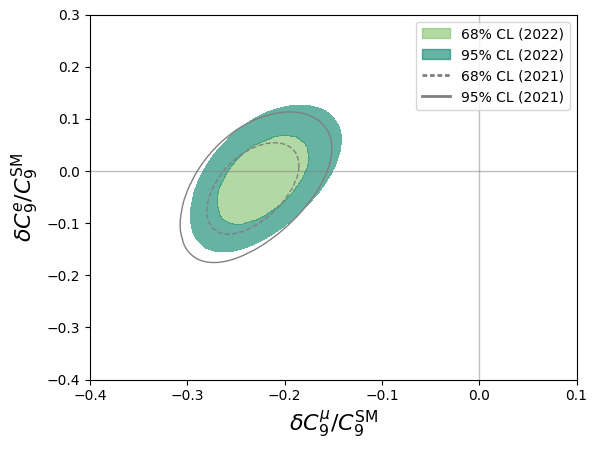}
\includegraphics[width=0.49\textwidth]{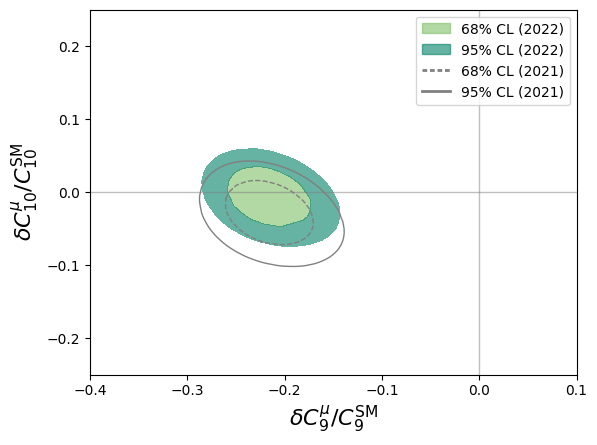}
\includegraphics[width=0.49\textwidth]{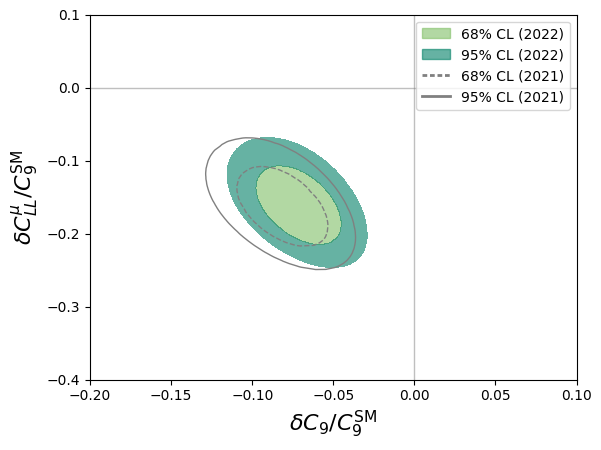}
\vspace{-0.3cm}
\caption{Two-dimensional fit to all rare $B$-decay observables.} 
\label{fig:all_2022}
\end{figure}

\begin{table}[b!]
\begin{center}
\setlength\extrarowheight{5pt}
\scalebox{0.75}{
\begin{tabular}{|C{1.2cm}|C{2.5cm}|C{1.7cm}|C{1.7cm}|}
\hline 
\multicolumn{4}{|c|}{All observables} \\[-4pt]										
\multicolumn{4}{|c|}{{\bf 2021 fit results}\quad ($\chi^2_{\rm SM}=225.8$)} \\ \hline			
& b.f. value & $\chi^2_{\rm min}$ & ${\rm Pull}_{\rm SM}$  \\										
\hline \hline										
$\delta C_{9} $    	& $ 	-0.99	\pm	0.13	 $ & $ 	186.2	 $ & $	6.3	\sigma	 $  \\
$\delta C_{9}^{e} $    	& $ 	0.79	\pm	0.20	 $ & $ 	207.7	 $ & $	4.3	\sigma	 $  \\
$\delta C_{9}^{\mu} $    	& $ 	-0.95	\pm	0.12	 $ & $ 	168.6	 $ & $	7.6	\sigma	 $  \\
\hline										
$\delta C_{10} $    	& $ 	0.32	\pm	0.18	 $ & $ 	222.3	 $ & $	1.9	\sigma	 $  \\
$\delta C_{10}^{e} $    	& $ 	-0.74	\pm	0.18	 $ & $ 	206.3	 $ & $	4.4	\sigma	 $  \\
$\delta C_{10}^{\mu} $    	& $ 	0.55	\pm	0.13	 $ & $ 	205.2	 $ & $	4.5	\sigma	 $  \\
\hline							          			
\hline 										
$\delta C_{\rm LL}^e$	& $ 	0.40	\pm	0.10	 $ & $ 	206.9	 $ & $	4.3	\sigma	 $  \\
$\delta C_{\rm LL}^\mu$            	& $ 	-0.49	\pm	0.08	 $ & $ 	180.5	 $ & $	6.7	\sigma	 $  \\
\hline
\end{tabular}
\qquad
\begin{tabular}{|C{1.2cm}|C{2.5cm}|C{1.7cm}|C{1.7cm}|}
\hline 
\multicolumn{4}{|c|}{All observables} \\[-4pt]										
\multicolumn{4}{|c|}{{\bf 2022 fit results}\quad ($\chi^2_{\rm SM}=253.5$)} \\ \hline			
& b.f. value & $\chi^2_{\rm min}$ & ${\rm Pull}_{\rm SM}$  \\										
\hline \hline										
$\delta C_{9} $    	& $ 	-0.95	\pm	0.13	 $ & $ 	215.8	 $ & $	6.1	\sigma	 $  \\
$\delta C_{9}^{e} $    	& $ 	0.82	\pm	0.19	 $ & $ 	232.4	 $ & $	4.6	\sigma	 $  \\
$\delta C_{9}^{\mu} $    	& $ 	-0.92	\pm	0.11	 $ & $ 	195.2	 $ & $	7.6	\sigma	 $  \\
\hline										
$\delta C_{10} $    	& $ 	0.08	\pm	0.16	 $ & $ 	253.2	 $ & $	0.5	\sigma	 $  \\
$\delta C_{10}^{e} $    	& $ 	-0.77	\pm	0.18	 $ & $ 	230.6	 $ & $	4.8	\sigma	 $  \\
$\delta C_{10}^{\mu} $    	& $ 	0.43	\pm	0.12	 $ & $ 	238.9	 $ & $	3.8	\sigma	 $  \\
\hline							          			
\hline										
$\delta C_{\rm LL}^e$	& $ 	0.42	\pm	0.10	 $ & $ 	231.4	 $ & $	4.7	\sigma	 $  \\
$\delta C_{\rm LL}^\mu$    	& $ 	-0.43	\pm	0.07	 $ & $ 	213.6	 $ & $	6.3	\sigma	 $  \\
\hline
\end{tabular}
} 
\caption{Comparison of the fits to all observables with the 2021 fit results on the left and the updated 2022 fits on the right.}
\label{tab:1D_full2022_all} 
\end{center} 
\end{table}
The 1-dim NP fits to the rare $B$-decays are given in Table~\ref{tab:1D_full2022_all}. 
As anticipated from the comparison of the fits to clean observables and the rest of the observables, the most favoured scenario to explain the overall data is lepton flavour violating NP in $\delta C_9^\mu$.
The other prominent scenarios are NP in $\delta C_{\rm LL}^\mu$ followed by lepton flavour universal NP in $\delta C_9$.
While the hierarchy of the favoured scenarios has not changed, it should be noted that NP in $\delta C_{10}^\mu$ is now less favoured which is mostly due to the updated combination for BR($B_s \to \mu^+ \mu^-$). This can also be seen in the decrease of the significance of $\delta C_{LL}^\mu$ compared to the 2021 fit results.
The decrease of preference of NP in $\delta C_{10}^\mu$ can also be seen in the 2-dim fits of Fig.~\ref{fig:all_2022}.

In the 1-~and 2-dim fits of Table~\ref{tab:1D_full2022_all} and Fig.~\ref{fig:all_2022} we have not shown the NP fits to the radiative coefficient $\delta C_7$, the scalar and pseudoscalar coefficients ($\delta C_{Q_{1,2}}$) or the coefficients where the hadronic currents are right-handed ($\delta C_i^{\prime}$) since they are all strongly constrained by data. The situation can in principle change when several coefficients can simultaneously contribute, this is clearly the case when doing a simultaneous fit to $\delta C_{10}^\mu$ and $\delta C_{Q_{1,2}}^\mu$~\cite{Arbey:2018ics} which would otherwise be severely constrained by BR($B_s\to\mu^+\mu^-$) if only one single coefficient would contribute.

\subsection{Multidimensional fit}
A multidimensional fit gives in principle a more realistic picture than assuming new physics contribution to only a single coefficient, as it is very unlikely for a UV-complete scenario to merely affect one coefficient while the rest of the coefficients are kept to their SM values. Therefore, here we consider a 20-dim fit varying all relevant Wilson coefficients (Table~\ref{tab:20D_full2022_all}).
Besides being more realistic, this multidimensional fit has the advantage of avoiding the look elsewhere effect (LEE) since LEE not only takes place when one makes a selected choice of observables but is also relevant in the case when a posteriori a subset of specific NP directions are assumed which is circumvented when all possible Wilson coefficients are varied. With a large set of free parameters and the limited decay modes there can be flat directions or non-sensitive NP coefficients that can be removed by considering the correlations and likelihood profiles in order to  get an ``effective'' number of degrees of freedom (dof$_{\rm eff}$). In the 20-dim fit we find degeneracy in $\delta C_{10}^e$ and $\delta C_{10}^{e\prime}$ which results in having dof$_{\rm eff}=19$.
With the current data, there are still several of the Wilson coefficients which are only loosely constrained, especially in the electron sector where there is less data.
The significance of the NP in our 20-dim fit is $5.5\sigma$, remaining the same as what we had found in Ref.~\cite{Hurth:2021nsi}.
\begin{table}[h!]
\begin{center}
\setlength\extrarowheight{5pt}
\scalebox{0.72}{
\begin{tabular}{|C{2.3cm}|C{2.3cm}|C{2.3cm}|C{2.3cm}|}
\hline																
\multicolumn{4}{|c|}{All observables  with $\chi^2_{\rm SM}=225.8,\;$ nr. obs.$=173$} \\											
\multicolumn{4}{|c|}{{\bf 2021 fit results}\quad ($\chi^2_{\rm min}=	 	151.6	;\; {\rm Pull}_{\rm SM}=	5.5(5.6)	\sigma$)} \\								
\hline \hline																
\multicolumn{2}{|c|}{$\delta C_7$} &  \multicolumn{2}{c|}{$\delta C_8$}\\																
\multicolumn{2}{|c|}{$	0.05	\pm	0.03	$} & \multicolumn{2}{c|}{$	-0.70	\pm	0.40	$}\\								
\hline																
\multicolumn{2}{|c|}{$\delta C_7^\prime$} &  \multicolumn{2}{c|}{$\delta C_8^\prime$}\\																
\multicolumn{2}{|c|}{$	-0.01	\pm	0.02	$} & \multicolumn{2}{c|}{$	0.00	\pm	0.80	$}\\								
\hline																
$\delta C_{9}^{\mu}$ & $\delta C_{9}^{e}$ & $\delta C_{10}^{\mu}$ & $\delta C_{10}^{e}$ \\																
$	-1.16	\pm	0.17	$ & $	-6.70	\pm	1.20	$ & $	0.20	\pm	0.21	$ &  \tiny{ degenerate w/} $\downarrow$ \\
\hline\hline																
$\delta C_{9}^{\prime \mu}$ & $\delta C_{9}^{\prime e}$ & $\delta C_{10}^{\prime \mu}$ & $\delta C_{10}^{\prime e}$ \\																
$	0.09	\pm	0.34	$ & $	1.90	\pm	1.50	$ & $	-0.12	\pm	0.20	$ & \tiny{ degenerate w/} $\uparrow$ \\
\hline\hline																
$\delta C_{Q_{1}}^{\mu}$ & $\delta C_{Q_{1}}^{e}$ & $\delta C_{Q_{2}}^{\mu}$ & $\delta C_{Q_{2}}^{e}$ \\																
$	0.04	\pm	0.10	$ & $	-1.50	\pm	1.50	$ & $	-0.09	\pm	0.10	$ & $	-4.10	\pm	1.5	$ \\
\hline\hline																
$\delta C_{Q_{1}}^{\prime \mu}$ & $\delta C_{Q_{1}}^{\prime e}$ & $\delta C_{Q_{2}}^{\prime \mu}$ & $\delta C_{Q_{2}}^{\prime e}$ \\																
$	0.15	\pm	0.10	$ & $	-1.70	\pm	1.20	$ & $	-0.14	\pm	0.11	$ & $	-4.20	\pm	1.2	$ \\
\hline																
\end{tabular}
\quad
\begin{tabular}{|C{2.3cm}|C{2.3cm}|C{2.3cm}|C{2.3cm}|}
\hline																
\multicolumn{4}{|c|}{All observables  with $\chi^2_{\rm SM}=253.5,\;$ nr. obs.$=183$} \\											
\multicolumn{4}{|c|}{{\bf 2022 fit results}\quad ($\chi^2_{\rm min}=	 	179.1	;\; {\rm Pull}_{\rm SM}=	5.5	(	5.5	)	\sigma$)} \\								
\hline \hline																
\multicolumn{2}{|c|}{$\delta C_7$} &  \multicolumn{2}{c|}{$\delta C_8$}\\																
\multicolumn{2}{|c|}{$	0.06	\pm	0.03	$} & \multicolumn{2}{c|}{$	-0.80	\pm	0.40	$}\\								
\hline																
\multicolumn{2}{|c|}{$\delta C_7^\prime$} &  \multicolumn{2}{c|}{$\delta C_8^\prime$}\\																
\multicolumn{2}{|c|}{$	-0.01	\pm	0.01	$} & \multicolumn{2}{c|}{$	-0.30	\pm	1.30	$}\\								
\hline																
$\delta C_{9}^{\mu}$ & $\delta C_{9}^{e}$ & $\delta C_{10}^{\mu}$ & $\delta C_{10}^{e}$ \\																
$	-1.14	\pm	0.19	$ & $	-6.50	\pm	1.90	$ & $	0.21	\pm	0.20	$ & \tiny{ degenerate w/} $\downarrow$ \\
\hline\hline																
$\delta C_{9}^{\prime \mu}$ & $\delta C_{9}^{\prime e}$ & $\delta C_{10}^{\prime \mu}$ & $\delta C_{10}^{\prime e}$ \\																
$	0.05	\pm	0.32	$ & $	1.40	\pm	2.30	$ & $	-0.03	\pm	0.19	$ & \tiny{ degenerate w/} $\uparrow$ \\
\hline\hline																
$\delta C_{Q_{1}}^{\mu}$ & $\delta C_{Q_{1}}^{e}$ & $\delta C_{Q_{2}}^{\mu}$ & $\delta C_{Q_{2}}^{e}$ \\																
$	0.04	\pm	0.20	$ & $	-1.60	\pm	1.70	$ & $	-0.15	\pm	0.08	$ & $	-4.10	\pm	0.9	$ \\
\hline\hline																
$\delta C_{Q_{1}}^{\prime \mu}$ & $\delta C_{Q_{1}}^{\prime e}$ & $\delta C_{Q_{2}}^{\prime \mu}$ & $\delta C_{Q_{2}}^{\prime e}$ \\																
$	-0.03	\pm	0.20	$ & $	-1.50	\pm	2.10	$ & $	-0.16	\pm	0.08	$ & $	-4.00	\pm	1.2	$ \\
\hline																
\end{tabular}
} 
\caption{Comparison of 20-dim fit to all observables with the 2022 (2021) result on the right (left). The Pull$_{\rm SM}$ in the parenthesis is given for  dof$_{\rm\, eff}=19$.}
\label{tab:20D_full2022_all} 
\end{center} 
\end{table}
%
%


\section{Conclusions}
We presented the 
NP fits to rare $B$ decays which include the recent measurements of $B_s \to \phi \mu^+ \mu^-$ observables and the lepton flavour violating ratios $R_{K^{*+}}$ and $R_{K_S}$ by LHCb. 
We have furthermore updated the BR($B_s \to \mu^+ \mu^-$) combination to include the very recent measurement by CMS. 
The main change in the NP fits is a reduction for the significance of a $\delta C_{10}^\mu$ solution or the scenarios involving it which is mainly due to the recent BR($B_s \to \mu^+ \mu^-$) measurement being in agreement with the SM value. However, the hierarchy of the favoured scenarios for the global fit has remained stable and the preferred scenario is still NP with $\delta C_9^\mu$. We also updated our twenty dimensional fit which avoids the look elsewhere effect finding a $5.5\sigma$ significance.

\providecommand{\href}[2]{#2}\begingroup\raggedright\endgroup
\end{document}